\documentclass[letterpaper]{article}
  \usepackage{aaai}
  \usepackage{times}
  \usepackage{helvet}
  \usepackage{courier}
  \usepackage[hyphens]{url}
  \usepackage{graphicx}
  \usepackage{graphicx}
\usepackage{xypic}
\usepackage{xy}
\usepackage{graphicx}
\begin{document}
\setcounter{secnumdepth}{2}
\title{Informal Data Transformation Considered Harmful}
\author{Eric Daimler, Ryan Wisnesky \\ Conexus AI}
\date{\today}
\newcommand{\LTO}{\sf}

\maketitle

\section*{}
\noindent
\hfill
\vspace{-.4in}
\includegraphics[width=3.315in]{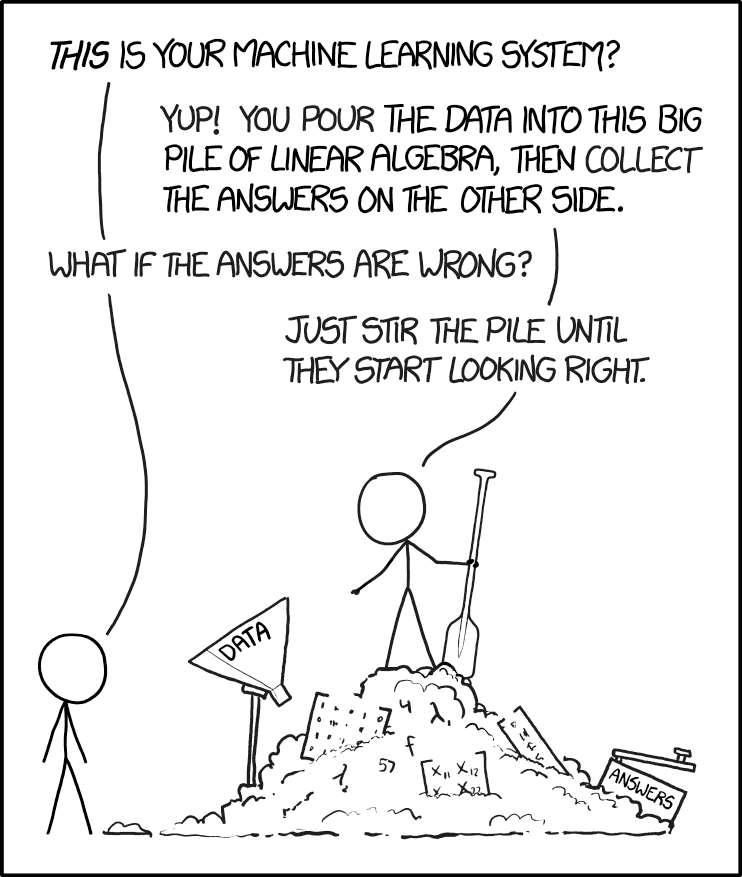}
Image used under a creative commons license; original available at \url{http://xkcd.com/1838/}.

\section{Introduction}

In this paper we take the common position~\cite{JSSv059i10} that AI systems are limited more by the integrity of the data they are learning from than the sophistication of their algorithms, and we take the uncommon position that the solution to achieving better data integrity in the enterprise is not to clean and validate data ex-post-facto whenever needed\footnote{the so-called ``data lake'' approach to data management, which can lead to data scientists spending 80\% of their time cleaning data~\cite{JSSv059i10}}, but rather to formally and automatically guarantee that data integrity is preserved as it transformed (migrated, integrated, composed, queried, viewed, etc) throughout the enterprise, so that data   and programs that depend on that data need not constantly be re-validated for every particular use. 
Computer scientists have been developing techniques for preserving data integrity during transformation since the 1970s~\cite{Doan:2012:PDI:2401764}; however, we agree with the authors of \cite{spencer} and others that these techniques are insufficient for the practice of AI and modern IT systems integration and we describe a modern mathematical approach based on {\it category theory}~\cite{BW,Awodey}, and the categorical query language CQL\footnote{ \url{http://categoricaldata.net}}, that is  sufficient for today's needs and also subsumes and unifies previous approaches.  

\subsection{Outline} 

To help motivate our approach, we next briefly summarize an application of CQL to a data science project undertaken jointly with  the Chemical Engineering department of Stanford University~\cite{kris}.
Then, in Section~\ref{sec:integrity} we review data integrity and in Section~\ref{sec:ct} we review category theory.  Finally, we describe the mathematics of our approach in Section~\ref{sec:fdm}, and conclude in Section~\ref{sec:conclusion}.  We present no new results, instead citing a line of work summarized in~\cite{wadt}.

\subsection{Motivating Case Study}
\label{sec:kris}

In scientific practice, computer simulation is now a third primary tool, alongside theory and experiment.  Within quantum materials engineering, {\it density functional theory (DFT)} calculations are invaluable for determining how electrons behave in materials, but  scientists typically do not share these calculations because they cannot guarantee that others will interpret them correctly, mitigating much of the value of simulation to begin with; for example, ease of replication.  Although there are many standardized formats for representing chemical structures, there are no such standards for more complicated entities such as the symmetry analysis of a chemical structure, the pseudo-potentials used in DFT calculation, density of states data resulting from a DFT calculation, and the DFT calculation itself.  Furthermore, many questions of interest depend not on a single calculation, but rather on ensembles of calculations, grouped in particular ways; for example, one is often interested in formation energies, i.e. a structure's energy relative to some reference energy, which depends on some arbitrarily-chosen mapping of its constituent elements to reference species, as well as the calculations for those reference species. 

The above descriptions represent a tiny fraction of the complexity of the systems computational scientists grapple with. In practice, scientists can only communicate structured raw data in tiny fragments (e.g. specific chemical structures) of the systems they try to model, which contain concepts at higher levels of abstraction such as chemical species, reaction mechanisms, and reaction networks, and the ability to freely exchange structured data at the level of abstraction which scientists actually work supports many scientific activities such as machine learning applications which thrive on large databases.

\begin{figure}
    \centering
    \includegraphics[width=3.3in]{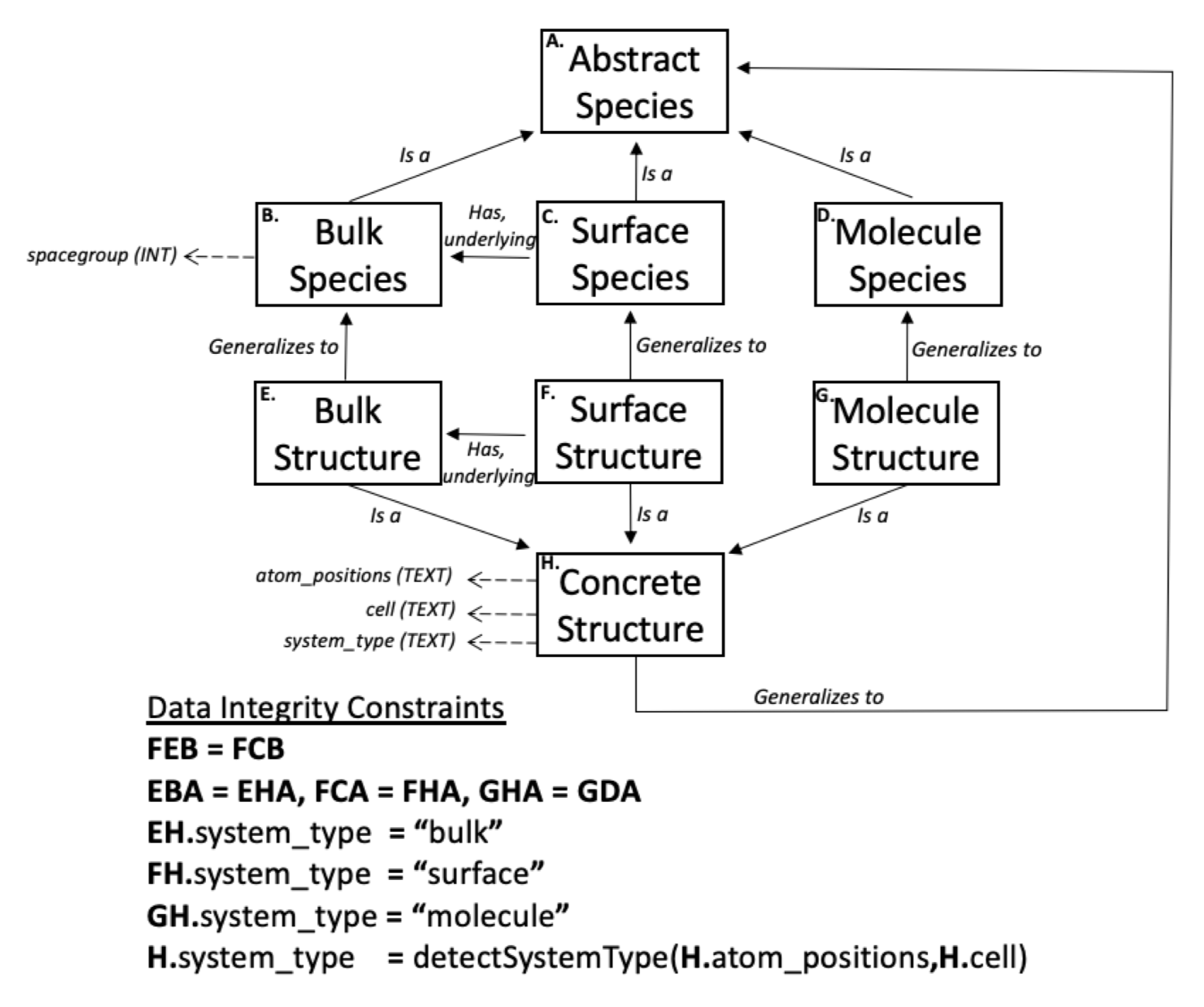}
    \caption{A Quantum Chemistry Schema}
    \label{fig:chem}
\end{figure}

The basic idea of the CQL solution to the above problem, which applies in many domains besides chemistry, is to design (or otherwise construct) a schema $C$ with a rich set of data integrity constraints that capture the properties that should be preserved upon transformation, and then to verify that any schema mappings out of $C$, for example by other scientists, respect these constraints, a process expedited by the CQL automated theorem prover, a technology we will describe later. In this way, data cannot be exported onto schemas that do not respect the  intentions of the original author.   An example schema for DFT calculations~\cite{kris} is shown in Figure~\ref{fig:chem}.  

%discuss examples from the literature where violations of data integrity have have led to catastrophic errors in AI systems, including both silent errors and errors that cannot be corrected for by downstream statistical methods even in principle (todo).  In the second half of this paper we review traditional approaches to ensuring data integrity during transformation, and then describe our approach and how it both subsumes existing approaches and is more broadly applicable.
\section{Data Integrity}
\label{sec:integrity}

By {\it data integrity}, we mean the conformance of a database to a collection of {\it data integrity constraints} expressed in some formal language, such as first-order logic.  When working with structured data, constraints are often built in to database {\it schemas}; for example, in SQL, a primary key constraint (stating for example that if two people have the same social security numbers, they must be the same person) can be given when defining the columns of a table.  Other example constraints include range constraints (that for example an age be $> 21$) and join decompositions (that for example state that a table is the join of two other tables; such denormalized data sets are common in data science, for performance reasons).  Another common constraint language is RDF/OWL~\cite{Doan:2012:PDI:2401764}. 

Data integrity is one of the mathematically quantifiable components of {\it data quality}, an informal and relative notion which roughly means that a database is ``fit'' for a particular purpose, such as supporting a particular forecasting model or building a particular data warehouse.  For this reason, computer scientists have long urged practitioners to, usually, use data integrity constraints to formalize as much of the concept of data quality as is possible in a given scenario, to provide both conceptual clarity and possibly be used in implementation.  However, unrestricted use of  constraints quickly leads to undecidable problems~\cite{Doan:2012:PDI:2401764}, as there is an innate trade-off between how expressive constraint languages are and how computationally difficult it is to reason about them.  Hence, mainstream data transformation tools (``ETL'' tools, such as Informatica PowerCenter and IBM DataStage) either do not support expressive constraints, or only use them internally.    

The failure of tooling described above is especially tragic because there is such a significant amount of theoretical work on applying constraints to data transformation, integration, cleaning, schema mapping, and more~\cite{Doan:2012:PDI:2401764}.  For example, it is possible to formally relate constraints to the random variables of statistical models and thereby demonstrate the absence of various statistical anomalies~\cite{JSSv059i10}.
It is doubly tragic because many practitioners, who are often not trained in computer science and many only be exposed to computer science through data transformation, may never be exposed to the idea that data quality can be formally and in many cases automatically enforced by using data integrity constraints. 

In the enterprise, data quality degradation due to loss of integrity during transformation is a systemic problem resistant to current techniques, and as such we propose to tackle it by building on top of a particular constraint language -- {\it category theory} -- that has become a de-facto constraint language for the practice of mathematics itself, in the sense that mathematicians use it to axiomatically define abstract structures such as groups using exactly the same kinds of data integrity constraints used in data management~\cite{Wells94sketches:outline}\footnote{A mathematician  would undoubtedly never refer to category theory as a constraint language, however.  Here we more precisely are referring to categorical logic, such as that of topoi.}.

\section{Category Theory}
\label{sec:ct}

Category theory~\cite{BW,Awodey} is the most recent branch of pure mathematics, originating in 1946 in algebraic topology.  There are three main concepts of study:  {\it categories}, {\it functors}, and {\it natural transformations}.

A {\it category} $\mathcal{C}$ consists of a set, ${\sf Ob}(\mathcal{C})$, the elements of which we call {\it objects}; a set ${\sf Mor}(\mathcal{C})$, the elements of which we call {\it morphisms}; functions ${\sf dom_\mathcal{C},cod_\mathcal{C}} : {\sf Mor}(\mathcal{C}) \to {\sf Ob}(\mathcal{C})$ called {\it domain} and {\it co-domain}; a function ${\sf id}_\mathcal{C} : {\sf Ob}(\mathcal{C}) \to {\sf Mor}(\mathcal{C})$; and a partial function $ \circ_\mathcal{C} : {\sf Mor}(\mathcal{C}) \times {\sf Mor}(\mathcal{C}) \to {\sf Mor}(\mathcal{C})$ called {\it composition} such that (where we omit the sub-scripted $\mathcal{C}$):

%\begin{footnotesize}
$$
{\sf dom}({\sf id}(o)) \ = \ {\sf cod}({\sf id}(o)) \ = \ o
$$
$$
{\sf cod}(g) = {\sf dom}(f) \ \Rightarrow \ {\sf dom}(f \circ g) = {\sf dom}(g) 
$$
$$
{\sf cod}(g) = {\sf dom}(f) \ \Rightarrow \ {\sf  \sf cod}(f \circ g) = {\sf cod}(f)
$$
$$
{\sf id}({\sf dom}(f)) \circ f \ = \ f \ =  \ {\sf id}({\sf cod}(f)) \circ f
$$
$$
( \ {\sf cod}(g) = {\sf dom}(f) \wedge {\sf cod}(h) = {\sf dom}(g) \Rightarrow 
$$
$$
(f \circ g) \circ h = f \circ (g \circ h) \ )
$$
%\end{footnotesize}
Note that the objects and morphisms of every category form a directed multi-graph.  An example category is {\sf Set}, the category whose objects are sets and whose morphisms are functions, with composition given by function application.    Another example is {\sf Group}, the category of groups and group homomorphisms.  Programming languages often form categories, with objects as types and morphisms as programs.

A {\it functor} $\mathcal{F} : \mathcal{C} \to \mathcal{D}$ from category $\mathcal{C}$ to category $\mathcal{D}$ consists of a function,  (also written) $\mathcal{F} : {\sf Ob}(\mathcal{C}) \to {\sf Ob}(\mathcal{D})$; and a function (also written) $\mathcal{F} : {\sf Mor}(\mathcal{C}) \to {\sf Mor}(\mathcal{D})$ that preserves identities and composition:
$$
{\sf dom}_\mathcal{D}(\mathcal{F}(f)) = \mathcal{F}({\sf dom}_\mathcal{C}(f)) \ \ \ \ \ \ \  \ \ 
\mathcal{F}({\sf id_\mathcal{C}}(o)) = {\sf id_\mathcal{D}}(\mathcal{F}(o)) 
$$
$$
{\sf cod}_\mathcal{D}(\mathcal{F}(f)) = \mathcal{F}({\sf cod}_\mathcal{C}(f))
 \ \ \ \ 
 \mathcal{F}(f \circ_\mathcal{C} g) = \mathcal{F}(f) \circ_\mathcal{D} \mathcal{G}(g)
$$
An example functor is the free group functor ${\sf free} : {\sf Set} \to {\sf Group}$ that takes each set to the free group generated by it, and each function to the associated unique group homomorphism.  Another example functor is the forgetful group functor ${\sf forget} : {\sf Group} \to {\sf Set}$ that takes each group to its underlying carrier set.  These two functors are not inverses, but are so called {\it adjoints}, a kind of generalization of the notion of inverse from which category theory derives much of its utility, but which we do not elaborate on here.

Finally, a {\it natural transformation} $\tau : \mathcal{F} \to \mathcal{G}$ between functors $\mathcal{F},\mathcal{G} : \mathcal{C} \to \mathcal{D}$ consists of, for every object $c \in {\sf Ob} (\mathcal{C})$ a morphism $\tau_c \in {\sf Mor}(\mathcal{D})$ such that
$$
{\sf dom}_\mathcal{D}(\tau_c) = \mathcal{F}(c)
 \ \ \ \ \ \ 
 {\sf cod}_\mathcal{D}(\tau_c) = \mathcal{G}(c) 
$$
$$
 \tau_{{\sf cod}_\mathcal{C}(f)} \circ_\mathcal{D} F(f) =
 \mathcal{G}(f) \circ_\mathcal{D} \tau_{{\sf dom}(f)}
$$
For every category $\mathcal{C}$, there is a category ${\sf Set}^\mathcal{C}$ whose objects are functors $\mathcal{C} \to {\sf Set}$ and whose morphisms are natural transformations.  ${\sf Set}^\mathcal{C}$ is a mathematical structure called a {\it topos} which can interpret first-order logic and set-theory, a fact~\cite{Wells94sketches:outline} that we will make use of in the next section.  Another fact we will use in the next section is that, ignoring minor issues relating to self-reference, there is a category whose objects are categories and whose morphisms are functors (and so in particular, functors compose, as do natural transformations).

\section{Functorial Data Migration}
\label{sec:fdm}

In this section we present our approach to formally and automatically ensuring that data integrity is preserved during transformation, comparing to previous approaches as we go along.  
%Because our approach generalizes most existing approaches, this section should be understandable to a high level by anyone with an understanding of the relational model of data as implemented in e.g., SQL.
Our key idea~\cite{wadt} is that database schemas are categories, and from that idea, an entire mathematical and algorithmic theory and practice of data transformation emerges, subsuming most current approaches such as SQL by virtue of the fact that category theory a kind of meta-theory for mathematics. 

\subsection{Algebraic Databases}

To use our approach, we start by defining a directed multi-graph to represent each database schema we are interested in.  The nodes of this graph are names for either types or database tables and the edges of this graph are names for either attributes or foreign keys, .  For example, if we are interested in a database about employees and departments, we may begin with the graph:

$$
\xymatrix{
{\sf Emp} \ar@(ul,ur)^{\sf mgr} \ar@/^/[rr]^{{\sf works}} \ar[rd]_{\sf name} & & {\sf Dept} \ar[ld]^{\sf name} \ar@/^/[ll]^{{\sf admin}} \\
& {\sf String} &
}
$$
For data integrity constraints, we use equational logic.  Continuing with our example, we might have:
$$
{\sf works}({\sf admin}(d)) = d 
$$
which express that every administrator works in the department he or she administers.  The graph and equations together determine a category whose objects are nodes in the graph and whose morphisms with domain $n_1$ and co-domain $n_2$ are (possibly empty) paths $n_1 \to n_2$ in the graph, where we identify paths that are equivalent under the equations.  
In this example, the induced category has infinitely many morphisms, because there are many paths through {\sf Emp} via {\sf mgr}.  

Having observed that database schemas are categories, we next observe that a database on schema $\mathcal{C}$ is a functor $\mathcal{C} \to {\sf Set}$.  Such a functor may presented as a set of tables, where we omit the infinite {\sf String} table that contains all strings:
$$
\begin{sf}
\begin{small}
\begin{tabular}{c|c|c|c}
{\sf Emp}&{\sf mgr }&{\sf works} & {\sf name} \\\hline 
101& 103 & q10 & Al  \\\hline 
102& 102 & x02 & Bob  \\\hline 
103& 103 & q10 & Carl  \\
\end{tabular} \ 
\begin{tabular}{c|c|c}
{\sf Dept}&{\sf admin}&{\sf name}\\\hline 
q10& 102 & CS \\\hline 
x02& 101 & Math \\ 
\end{tabular}
\end{small}
\end{sf}
$$
It is easy to see that these tables satisfy the data integrity constraints.  Indeed, when $\mathcal{C}$ is presented by generating morphisms and equations, every functor $\mathcal{C} \to {\sf Set}$ will satisfy the equations of $\mathcal{C}$, a useful property of our formalism.  We will not illustrate natural transformations between two databases, but they  correspond precisely to their relational counterparts~\cite{Doan:2012:PDI:2401764}.

It is important to note that although we are using tables in the example above, our approach is not limited to relational data; for example, we may just as easily use a representation of databases as graphs, such as found in a triple store:
$$
\xymatrix{
\LTO{CS} &\ar[l]^{\sf name} \LTO{q10} & & \ar[ll]^{\sf works} \LTO{101} \ar[r]^{\sf first} \ar@/_1.2pc/[dd]_(.3){\sf mgr} & \LTO{Al}  \\
\LTO{Math} &\ar[l]^{\sf name} \LTO{x02} & & \ar[ll]^(.4){\sf works} \LTO{102} \ar[r]^{\sf first}  \ar@(r,d)[]^{\sf mgr} & \LTO{Bob}  \\
& & & \ar@/^1.2pc/[lluu]^(.3){\sf works} \LTO{103} \ar@(r,d)[]^{\sf mgr} \ar[r]^{\sf first}  & \LTO{Carl}  \\
}
%\vspace{-.1in}
$$

\subsection{Data Transformation}
 Given a functor $\mathcal{F} : \mathcal{C} \to \mathcal{D}$, we can convert a database $\mathcal{J}$ on schema $\mathcal{D}$ to a database $\Delta_\mathcal{F}(\mathcal{J})$ on schema $\mathcal{C}$ by composition pre-composition with $\mathcal{F}$:
$$
\xymatrix{
\mathcal{C} \ar[r]^{\mathcal{F}} \ar@/_1pc/[rr]_{\Delta_\mathcal{F}(\mathcal{J}) : \mathcal{C} \to {\sf Set} \ := \ \mathcal{J} \circ \mathcal{F}} & \mathcal{D} \ar[r]^{\mathcal{J}} & {\sf Set} \\
}
$$
The operation (in fact, functor) $\Delta_F$ that takes databases on schema $\mathcal{D}$ and converts them to schema $\mathcal{C}$ does not (in general) admit an inverse, but it does (always) admit two inverse-like operations, $\Sigma_F$ and $\Pi_F$, called the left and right {\it adjoints} of $\Delta_F$, respectively, that convert databases on schema $\mathcal{C}$ to schema $\mathcal{D}$; i.e., they migrate data in the opposite direction as $\Delta_F$.  The operations $\Delta_F,\Sigma_F,\Pi_F$ form a core part of our approach are are used many times over in a typical CQL program.

In our approach it is natural to call a functor $\mathcal{F} : \mathcal{C} \to \mathcal{D}$ a ``schema mapping'', because functors are called maps between categories.  However, in relational database theory, there are not three data migration operations corresponding to one schema mapping, there is only one, with a semantics (called ``chase semantics''~\cite{Doan:2012:PDI:2401764}) broadly similar to our $\Sigma_F$.  Our approach thus generalizes relational schema mappings~\cite{Doan:2012:PDI:2401764}.  Additionally, pairs of schema mappings with a common co-domain generalize SQL's SELECT-FROM-WHERE queries, but without issues relating to NULLs and without the need to join columns that are connected by foreign keys.  Finally, our approach generalizes~\cite{wadt} to encompass all of the classes of data integrity constraints studied in database theory, such as existential horn-clause logic~\cite{Doan:2012:PDI:2401764}.  In short, there is little to be lost in moving to a category-theoretic (``functorial'') approach to data migration from the traditional, relational approach.  

In fact, there is much to be gained as well: because functors preserve equations, as discussed earlier, data migrations in our approach also always preserve data integrity constraints, meaning that queries ``cannot go wrong''.  The price to be paid for this compile-time assurance is that checking if a functor $\mathcal{F} : \mathcal{C} \to \mathcal{D}$ is actually functorial and not just an assignment of objects to objects and arrows to arrows is an undecidable problem.  The reason is that, when $\mathcal{C}$ and $\mathcal{D}$ are presented by generating morphisms and equations, we must check, for each equation $p = q$ in $\mathcal{C}$, that $\mathcal{F}(p) = \mathcal{F}(q)$ in $\mathcal{D}$, which is not decidable in general~\cite{Baader:1998:TR:280474}.  Hence, for our approach to be feasible in practice, automated theorem proving techniques~\cite{Baader:1998:TR:280474} must be used, and their development is one of the key contributions of the open-source CQL query language, which implements the $\Delta,\Sigma,\Pi$ data migrations, and many other constructions, in software.

\section{Conclusion}
\label{sec:conclusion}

In this paper we have presented a vision for a formally verified data transformation infrastructure, based on the mathematics of category theory and the categorical query language CQL, and described a motivating case study from data science.  Our approach to data migration and integration extends existing approaches from relational database theory, while providing an extended semantics that has already proved indispensable in many projects, including~\cite{sub},\cite{sub2}, and~\cite{kris}.  Given that data quality is the primary obstacle in unlocking the power of AI, by transitivity we expect our approach to data transformation to be fundamental to unlocking the power of AI. 

%Work on our approach continues.  

\begin{footnotesize}
\bibliographystyle{aaai}
\bibliography{bib}
\end{footnotesize}

%\end{multicols*}

\end{document}